\documentclass{article}
\usepackage{amsmath}
\usepackage{mathtools}
\usepackage{amsfonts, bbm, amsmath, float, courier}
\usepackage{authblk}

\input{tcilatex}
\begin{document}

\title{Estimation of Dynamic Panel Threshold Model using Stata\thanks{%
This work was supported by the Ministry of Education of the Republic of
Korea and the National Research Foundation of Korea (NRF-2017S1A5A8019707).
Seo acknowledges financial support from the center for national
competitiveness in the institute of economic research.}}

\author[a]{Myung Hwan Seo}
\author[b]{Sueyoul Kim}
\author[c]{Young-Joo Kim}

\affil[a]{\footnotesize Department of Economics, Seoul National University} %
\affil[b]{\footnotesize Department of Economics, University of Maryland}
\affil[c]{\footnotesize Department of Economics, Hongik University}

\maketitle

\begin{abstract}
We develop a Stata command \texttt{xthenreg} to implement the
first-differenced GMM estimation of the dynamic panel threshold model, which
Seo and Shin (2016, \textit{Journal of Econometrics }195: 169-186) have
proposed. Furthermore, We derive the asymptotic variance formula for a kink
constrained GMM estimator of the dynamic threshold model and include an
estimation algorithm. We also propose a fast bootstrap algorithm to
implement the bootstrap for the linearity test. The use of the command is
illustrated through a Monte Carlo simulation and an economic application.
\end{abstract}

\affil[a]{Department of Economics, Seoul National University}

\affil[b]{Department of Economics, Hongik University}

\section{Introduction}

The panel model with threshold effects in Hansen (1999) has been widely
used in the empirical research. Hansen's fixed effect estimator has been
applied to applications on the investment decision of firms under financial
constraints, the relation between fiscal deficit and economic growth (Adam
and Bevan 2005), inflation and growth (Khan and Ssnhadji 2001) and others.
The threshold effect in the model allows for the asymmetric effect of the
exogeneous variables depending on whether the threshold variable is above or
below the unknown threshold. The threshold variable is typically dictated by
the economic model. For instance, in the investment decision problem the
size of the firm is often considered as a candidate threshold variable. 
Wang (2015) has developed Stata command \texttt{xthreg} to compute Hansen's estimator.

Hansen's (1999) model is static and his fixed effect estimator requires the
covariates to be strongly exogeneous for the estimator to be consistent.
However, the strong exogeneity can be restrictive in many real applications.
Thus, the model has been extended to the dynamic panel model with a
potentially endogenous threshold variable by Seo and Shin (2016). Their
model allows for the lagged dependent variables and endogeneous covariates.
Indeed, various applications of Hansen's fixed effect estimation can benefit
from dynamic modeling. For instance, the investment decision depends on the
previous period's investment and the panel threshold autoregressive model is
another example of dynamic models.

We develop Stata commands for the first-differenced generalized method of
moments (GMM) estimators and the associated asymptotic variance estimator
that are proposed by Seo and Shin (2016) as well as the linearity testing to
test for the presence of a threshold effect. While the previous command 
\texttt{xthreg} computes the fixed-effect estimator and thus it is not
consistent under this general setting, our command \texttt{xthenreg}
produces a consistent and asymptotically normal estimates.

In addition, we propose a computationally more attractive bootstrap
algorithm to implement the linearity test than the
nonparametric i.i.d. bootstrap that is originally proposed by Seo and Shin
(2016). Furthermore, we present a constrained GMM estimator that reflects
the kink restriction that has become more popular recent years, as in e.g.
Zhang et al. (2017), along with its asymptotic variance formula and a
consistent estimator.

The paper is organized as follows. Section 2 introduces the dynamic
threshold panel model and the first-differenced GMM estimator. It also
presents the asymptotic variance formula for a kink constrained estimator
and a bootstrap algorithm for the linearity test. Section 3 explains the
command \texttt{xthenreg}. Its use is illustrated in Section 4 and 5 through
Monte Carlo simulations and an application. Section 6 concludes.

\section{Model}

The dynamic panel threshold model is given by%
\begin{equation*}
y_{it}=x_{it}^{\prime }\beta +\left( 1,x_{it}^{\prime }\right) \delta
1\left\{ q_{it}>\gamma \right\} +\mu _{i}+\varepsilon
_{it},\;i=1,...,n;\;t=1,...,T,
\end{equation*}%
where $x_{it}$ may include lagged dependent variables and $q_{it}$ is the
threshold variable. We assume $T$ is fixed while the sample size $n$ grows
to infinity. Thus, we remove the incidental parameter $\mu _{i}$ by the
first difference transformation and estimate the unknown parameters $\theta
=\left( \beta ^{\prime },\delta ^{\prime },\gamma \right) ^{\prime }$
through the GMM. The following describes the GMM method as in Seo and Shin
(2016).

Specifically, set an $l$-dimensional vector of instrument variables, $\left(
z_{it_{0}}^{\prime },....,z_{iT}^{\prime }\right) ^{\prime }$ from the
lagged variables and exogenous variables, where $2<t_{0}\leq T.$ Next,
construct the sample moment 
\begin{equation*}
\bar{g}_{n}\left( \theta \right) =\bar{g}_{1n}-\bar{g}_{2n}\left( \gamma
\right) \left( \beta ^{\prime },\delta ^{\prime }\right) ^{\prime }=\frac{1}{%
n}\sum_{i=1}^{n}g_{1i}-\frac{1}{n}\sum_{i=1}^{n}g_{2i}\left( \gamma \right)
\left( \beta ^{\prime },\delta ^{\prime }\right) ^{\prime },
\end{equation*}%
where%
\begin{equation}
g_{1i}=\left( 
\begin{array}{c}
z_{it_{0}}\Delta y_{it_{0}} \\ 
\vdots \\ 
z_{iT}\Delta y_{iT}%
\end{array}%
\right) ,\ g_{2i}\left( \gamma \right) =\left( 
\begin{array}{c}
z_{it_{0}}\left( \Delta x_{it_{0}}^{\prime },\mathbf{1}_{it_{0}}\left(
\gamma \right) ^{\prime }X_{it_{0}}\right) \\ 
\vdots \\ 
z_{iT}\left( \Delta x_{iT}^{\prime },\mathbf{1}_{iT}\left( \gamma \right)
^{\prime }X_{iT}\right)%
\end{array}%
\right) ,  \label{g_i}
\end{equation}%
with $\Delta $ signifying the first difference operator and 
\begin{equation*}
X_{it}=\left( 
\begin{array}{c}
\left( 1,x_{it}^{\prime }\right) \\ 
\left( 1,x_{i,t-1}^{\prime }\right)%
\end{array}%
\right) \ \ \text{and\ \ \ }\underset{2\times 1}{\mathbf{1}_{it}\left(
\gamma \right) }=\left( 
\begin{array}{c}
1\left\{ q_{it}>\gamma \right\} \\ 
-1\left\{ q_{it-1}>\gamma \right\}%
\end{array}%
\right) .
\end{equation*}%
Then, introduce the GMM criterion function with a weight matrix $W_{n},$%
\begin{equation}
\bar{J}_{n}\left( \theta \right) =\bar{g}_{n}\left( \theta \right) ^{\prime
}W_{n}\bar{g}_{n}\left( \theta \right) ,  \label{Jn}
\end{equation}%
which is minimized to produce a GMM estimate $\hat{\theta}$.

The minimization is done by the grid search since for each fixed $\gamma$
the model becomes the linear panel with a fixed effect, which yields the
closed-form solution 
\begin{equation}
\left( \hat{\beta}\left( \gamma \right) ^{\prime },\hat{\delta}\left( \gamma
\right) ^{\prime }\right) ^{\prime }=\left( \bar{g}_{2n}\left( \gamma
\right) ^{\prime }W_{n}~\bar{g}_{2n}\left( \gamma \right) \right) ^{-1}\bar{g%
}_{2n}\left( \gamma \right) ^{\prime }W_{n}~\bar{g}_{1n}  \label{eq:delta}
\end{equation}%
and the criterion function $\bar{J}_{n}\left( \theta \right) $ is a step
function over $\gamma $ with at most $nT$ jumps. However, it is worthwhile
to note that this algorithm is different from splitting the sample into two
and applying the linear GMM for each partitioned sample.

For the weight matrix, either $W_{n}=I_{l}$ or 
\begin{equation}
W_{n}=\left( 
\begin{array}{cccc}
\frac{2}{n}\sum_{i=1}^{n}z_{it_{0}}z_{it_{0}}^{\prime } & \frac{-1}{n}%
\sum_{i=1}^{n}z_{it_{0}}z_{it_{0}+1}^{\prime } & 0 & \cdots \\ 
\frac{-1}{n}\sum_{i=1}^{n}z_{it_{0}+1}z_{it_{0}}^{\prime } & \frac{2}{n}%
\sum_{i=1}^{n}z_{it_{0}+1}z_{it_{0}+1}^{\prime } & \ddots & \ddots \\ 
0 & \ddots & \ddots & \frac{-1}{n}\sum_{i=1}^{n}z_{iT-1}z_{iT}^{\prime } \\ 
\vdots & \ddots & \frac{-1}{n}\sum_{i=1}^{n}z_{iT}z_{iT-1}^{\prime } & \frac{%
2}{n}\sum_{i=1}^{n}z_{iT}z_{iT}^{\prime }%
\end{array}%
\right) ^{-1}  \label{wn0}
\end{equation}%
was proposed in the first step and it is updated to 
\begin{equation}
W_{n}=\left( \frac{1}{n}\sum_{i=1}^{n}\hat{g}_{i}\hat{g}_{i}^{\prime }-\frac{%
1}{n^{2}}\sum_{i=1}^{n}\hat{g}_{i}\sum_{i=1}^{n}\hat{g}_{i}^{\prime }\right)
^{-1},  \label{wn}
\end{equation}%
where $\hat{g}_{i}=\left( \widehat{\Delta \varepsilon }_{it_{0}}z_{it_{0}}^{%
\prime },...,\widehat{\Delta \varepsilon }_{iT}z_{iT}^{\prime }\right)
^{\prime }$ and $\widehat{\Delta \varepsilon }_{it}$ is the residual from
the first step estimation.

It was shown by Seo and Shin (2016) that under suitable regularity conditions%
\footnote{%
One of the conditions allows for $\delta _{0}$ to be both fixed and shrinking toward zero at $%
n^{-\alpha }.$} the GMM estimator is asymptotically normal. Specifically,  
\begin{equation*}
\left( 
\begin{array}{c}
\sqrt{n}\left( 
\begin{array}{c}
\hat{\beta}-\beta _{0} \\ 
\hat{\delta}-\delta _{n}%
\end{array}%
\right) \\ 
n^{1/2-\alpha }\left( \hat{\gamma}-\gamma _{0}\right)%
\end{array}%
\right) \overset{d}{\longrightarrow }\mathcal{N}\left( 0,\left( G^{\prime
}\Omega ^{-1}G\right) ^{-1}\right) ,
\end{equation*}%
where $G=\left( G_{\beta },G_{\delta }\left( \gamma _{0}\right) ,G_{\gamma
}\left( \gamma _{0}\right) \right) $ with 
\begin{equation*}
\underset{l\times k_{1}}{G_{\beta }}=\left[ 
\begin{array}{c}
-\mathrm{E}\left( z_{it_{0}}\Delta x_{it_{0}}^{\prime }\right) \\ 
\vdots \\ 
-\mathrm{E}\left( z_{iT}\Delta x_{iT}^{\prime }\right)%
\end{array}%
\right] ,\ \ \ \underset{l\times \left( k_{1}+1\right) }{G_{\delta }}\left(
\gamma \right) =\left[ 
\begin{array}{c}
-\mathrm{E}\left( z_{it_{0}}\mathbf{1}_{it_{0}}\left( \gamma \right)
^{\prime }X_{it_{0}}\right) \\ 
\vdots \\ 
-\mathrm{E}\left( z_{iT}\mathbf{1}_{iT}\left( \gamma \right) ^{\prime
}X_{iT}\right)%
\end{array}%
\right] ,
\end{equation*}%
and 
\begin{equation*}
\underset{l\times 1}{G_{\gamma }}\left( \gamma \right) =\left[ 
\begin{array}{c}
\left\{ \mathrm{E}_{t_{0}-1}\left[ z_{it_{0}}\left( 1,x_{it_{0}-1}^{\prime
}\right) |\gamma \right] p_{t_{0}-1}\left( \gamma \right) -\mathrm{E}_{t_{0}}%
\left[ z_{it_{0}}\left( 1,x_{it_{0}}^{\prime }\right) |\gamma \right]
p_{t_{0}}\left( \gamma \right) \right\} \delta _{0} \\ 
\vdots \\ 
\left\{ \mathrm{E}_{T-1}\left[ z_{iT}\left( 1,x_{iT-1}^{\prime }\right)
|\gamma \right] p_{T-1}\left( \gamma \right) -\mathrm{E}_{T}\left[
z_{iT}\left( 1,x_{iT}^{\prime }\right) |\gamma \right] p_{T}\left( \gamma
\right) \right\} \delta _{0}%
\end{array}%
\right] ,
\end{equation*}%
where $\mathrm{E}_{t}\left[ \cdot |\gamma \right] $ denotes the conditional
expectation given $q_{it}=\gamma $ and $p_{t}\left( \cdot \right) $ denotes
the density of $q_{it}.$

The estimation of the asymptotic variance is standard, that is, 
\begin{equation*}
\hat{\Omega}\left( \theta \right) =\frac{1}{n}\sum_{i=1}^{n}g_{i}\left(
\theta \right) g_{i}\left( \theta \right) ^{\prime }-\frac{1}{n}%
\sum_{i=1}^{n}g_{i}\left( \theta \right) \frac{1}{n}\sum_{i=1}^{n}g_{i}%
\left( \theta \right) ^{\prime },
\end{equation*}%
where $g_{i}\left( \theta \right) =g_{1i}+g_{2i}\left( \gamma \right) \left(
\beta ^{\prime },\delta ^{\prime }\right) ^{\prime }$, and 
\begin{equation*}
\hat{G}_{\beta }=\left[ 
\begin{array}{c}
-\frac{1}{n}\sum_{i=1}^{n}z_{it_{0}}\Delta x_{it_{0}}^{\prime } \\ 
\vdots \\ 
-\frac{1}{n}\sum_{i=1}^{n}z_{iT}\Delta x_{iT}^{\prime }%
\end{array}%
\right] ,\ \ \ \hat{G}_{\delta }\left( \gamma \right) =\left[ 
\begin{array}{c}
-\frac{1}{n}\sum_{i=1}^{n}z_{it_{0}}\mathbf{1}_{it_{0}}\left( \gamma \right)
^{\prime }X_{it_{0}} \\ 
\vdots \\ 
-\frac{1}{n}\sum_{i=1}^{n}z_{iT}\mathbf{1}_{iT}\left( \gamma \right)
^{\prime }X_{iT}%
\end{array}%
\right]
\end{equation*}%
\begin{equation}
\hat{G}_{\gamma }\left( \theta \right) =\left[ 
\begin{array}{c}
\frac{1}{nh}\sum_{i=1}^{n}z_{it_{0}}\left[ \left( 1,x_{it_{0}-1}^{\prime
}\right) ^{\prime }K\left( \frac{\gamma -q_{it_{0}-1}}{h}\right) -\left(
1,x_{it_{0}}^{\prime }\right) ^{\prime }K\left( \frac{\gamma -q_{it_{0}}}{h}%
\right) \right] \delta \\ 
\vdots \\ 
\frac{1}{nh}\sum_{i=1}^{n}z_{iT}\left[ \left( 1,x_{iT-1}^{\prime }\right)
^{\prime }K\left( \frac{\gamma -q_{iT-1}}{h}\right) -\left( 1,x_{iT}^{\prime
}\right) ^{\prime }K\left( \frac{\gamma -q_{iT}}{h}\right) \right] \delta%
\end{array}%
\right] ,  \label{Ghatgm}
\end{equation}%
which is the Nadaraya-Watson kernel estimator for some kernel $K$ and
bandwidth $h$ such as the Gaussian kernel and Silverman's rule of thumb. We
plug in $\theta =\hat{\theta}$.

\subsection{Kink Model}

Although the threshold model typically implies the presence of a
discontinuity of the regression function, it may mean the presence of a kink
not a jump if $\left( 1,x_{it}^{\prime }\right) \delta =\kappa \left(
q_{it}-\gamma \right) $ for some $\kappa $. It happens when one element of $%
x_{it}$ is $q_{it}$ with the coefficient $\kappa $ and the first element of $%
\delta $ equals to $-\gamma \kappa .$ Under these restrictions, the model
becomes 
\begin{equation*}
y_{it}=x_{it}^{\prime }\beta +\kappa \left( q_{it}-\gamma \right) 1\left\{
q_{it}>\gamma \right\} +\alpha _{i}+\varepsilon
_{it},\;i=1,...,n;\;t=1,...,T.
\end{equation*}

Even when the true model is a kink one, it is shown that the asymptotic
distribution of the GMM estimator in the preceding section is valid. This is
in contrast to the least squares estimator for the linear regression, for
which Hidalgo et al. (2019) have shown that the cube root phenomenon appears.

The asymptotic distribution of the constrained GMM estimator of $(\beta
,\kappa ,\gamma )$ that imposes the kink restriction can also be derived for
the same reasoning as in Seo and Shin (2016). Specifically, the asymptotic
variance is given by redefining $G=\left( G_{\beta },G_{\kappa },G_{\gamma
}\right) ,$ where $G_{\beta }$ is the same as above and 
\begin{eqnarray*}
G_{\kappa } &=&\left( 
\begin{array}{c}
\mathrm{E}z_{it_{0}}\left( (q_{it_{0}}-\gamma _{0})1\{q_{it_{0}}>\gamma
_{0}\}-(q_{i,t_{0}-1}-\gamma _{0})1\{q_{i,t_{0}-1}>\gamma _{0}\}\right) \\ 
\vdots \\ 
\mathrm{E}z_{iT}\left( (q_{iT}-\gamma _{0})1\{q_{iT}>\gamma
_{0}\}-(q_{i,T-1}-\gamma _{0})1\{q_{i,T-1}>\gamma _{0}\}\right)%
\end{array}%
\right) , \\
G_{\gamma } &=&\kappa _{0}\left( 
\begin{array}{c}
\mathrm{E}z_{it_{0}}\left( 1\{q_{i,t_{0}-1}>\gamma
_{0}\}-1\{q_{it_{0}}>\gamma _{0}\}\right) \\ 
\vdots \\ 
\mathrm{E}z_{iT}\left( 1\{q_{i,T-1}>\gamma _{0}\}-1\{q_{iT}>\gamma
_{0}\}\right)%
\end{array}%
\right) .
\end{eqnarray*}%
The estimation of these terms is analogous to that of $G_{\delta }$ and $%
G_{\gamma }$ in the preceding section.

\subsection{Bootstrap Test of Linearity}

\label{Sec:Bootstrap}This section proposes a fast bootstrap algorithm to
test for the presence of the threshold effect, that is, the null hypothesis 
\begin{equation}
\mathcal{H}_{0}:\delta _{0}=0,\ \ \text{for any }\gamma \in \Gamma ,
\label{H0}
\end{equation}%
where $\Gamma $ denotes the parameter space for $\gamma $, against the
alternative hypothesis 
\begin{equation*}
\mathcal{H}_{1}:\delta _{0}\neq 0,\ \ \text{for some }\gamma \in \Gamma .
\end{equation*}

A standard approach is to employ a supremum type statistic to take care of
the loss of identification under the null, that is, 
\begin{equation*}
\text{supW}=\sup_{\gamma \in \Gamma }\mathcal{W}_{n}\left( \gamma \right) ,
\end{equation*}%
where $\mathcal{W}_{n}\left( \gamma \right) $ is the standard Wald statistic
for each fixed $\gamma $, that is, 
\begin{equation}
\mathcal{W}_{n}\left( \gamma \right) =n\hat{\delta}\left( \gamma \right)
^{\prime }\hat{\Sigma}_{\delta }\left( \gamma \right) ^{-1}\hat{\delta}%
\left( \gamma \right) ,  \label{eq:wng}
\end{equation}%
where $\hat{\delta}\left( \gamma \right) $ is the GMM estimator of $\delta $
for a given $\gamma $, and 
\begin{equation*}
\hat{\Sigma}_{\delta }\left( \gamma \right) =R\left( \hat{V}_{s}\left(
\gamma \right) ^{\prime }\hat{V}_{s}\left( \gamma \right) \right)
^{-1}R^{\prime },
\end{equation*}%
a consistent asymptotic variance estimator, where $R=\left( \mathbf{0}%
_{\left( k_{1}+1\right) \times k_{1}}\mathbf{,}I_{k_{1}+1}\right) ,$ and $%
\hat{V}_{s}\left( \gamma \right) =\hat{\Omega}\left( \hat{\theta}\left(
\gamma \right) \right) ^{-1/2}\left( \hat{G}_{\beta },\hat{G}_{\delta
}\left( \hat{\theta}\left( \gamma \right) \right) \right) $.

Since the asymptotic distribution is not pivotal, we propose a bootstrap
algorithm, which is faster than the i.i.d. bootstrap proposed in Seo and
Shin (2016). Specifically,

\begin{enumerate}
\item Draw $\left\{ \eta _{i}\right\} _{i=1}^{n}$ independently from the
standard normal.

\item Recall the definition of $\hat{\delta}\left( \gamma \right) $ in (\ref%
{eq:delta}) and compute $\hat{\delta}\left( \gamma \right) ^{\ast }$ by
replacing $\Delta y_{it}$ with $\Delta y_{it}^{\ast }=\widehat{\Delta
\varepsilon _{it}}\eta _{i}$, where $\widehat{\Delta \varepsilon _{it}}%
=\Delta y_{it}-\Delta x_{it}^{\prime }\hat{\beta}-\hat{\delta}^{\prime
}X_{it}^{\prime }\mathbf{1}_{it}\left( \hat{\gamma}\right) $ is the residual
from the original sample.

\item Compute a bootstrap statistic $\mathcal{W}_{n}^{\ast }(\gamma )=n\hat{%
\delta}\left( \gamma \right) ^{\ast \prime }\hat{\Sigma}_{\delta }\left(
\gamma \right) ^{-1}\hat{\delta}\left( \gamma \right) ^{\ast }$ and its
supremum over $\Gamma $ to get $\text{supW}^{\ast }$.

\item Repeat step 1-3 $B$ times and compute the empirical proportion of supW$%
^{\ast }$ bigger than supW.
\end{enumerate}

\section{Command}
\subsection{Syntax}
\hangindent=1em
\texttt{xthenreg} $depvar$ $indepvars$  [$if$] [$in$] \newline 
[,  \texttt{\underline{endo}genous}($varlist$) \texttt{inst($varlist$)} \texttt{kink} \texttt{static} \newline
\texttt{grid\_num}($integer$ 20) \texttt{trim\_rate}($real$ 0.4) \texttt{h\_0}($real$ 1.5) \texttt{boost}($real$ 0)]\\

where $depvar$ is the dependent variable and $indepvars$ are the independent variables. There are several comments for users.
\begin{enumerate}
\item \texttt{xtset} should be done before running this. Moreover variables must be sorted by (i) panel variable and (ii) time variable beforehand.
\item Strongly balanced panel data is required.
\item Inputs should be put as \texttt{y q x1 x2 $\cdots$}, where \texttt{q} is the threshold variable and \texttt{x1 x2 $\cdots$} are other independent variables.
\item \texttt{moremata} library is required since this command use \texttt{mm\_quantile} function.
\item When there are endogeneous independent variables, $\texttt{\underline{endo}genous}$ option should be set. For example, if \texttt{x1} is exogeneous and \texttt{x2} is endogeneous, the input must be \texttt{y q x1, endo(x2)}.
\end{enumerate}

\subsection{Options}
\texttt{endogenous}($varlist$)      specifies endogeneous independent variables. The endogeneous variables must be excluded from the list of independent variables before the comma.\\\\
\texttt{inst}($varlist$) specifies the list of additional instrumental variables. \\\\
\texttt{static}  sets the model static. The default model is dynamic.  In contrast with dynamic model, static model does not automatically include \texttt{L.y} as independent variable. \\\\
\texttt{kink} sets the model kink. \\\\
\texttt{grid\_num}($integer$) determines the number of grid points to estimate the threshold $\gamma$. The default is 20.\\\\
\texttt{trim\_rate}($real$) determines the trim rate when constructing a grid for estimating r. The default is 0.4.\\\\
\texttt{h\_0}($real$) determines a parameter for Silverman's rule of thumb used to kernel estimation.  The default is 1.5.\\\\
\texttt{boost}($integer$) The number of bootstrapping for linearity test. The default is 0.\\

\subsection{Stored Results}

\texttt{xthenreg} stores the following results in \texttt{e( )}:\\\\
Scalars
\begin{itemize}
\item[] \texttt{e(N)}	The number of units of panel data
\item[] \texttt{e(T)}	The time length of panel data
\item[] \texttt{e(boots\_p)}	$p\textit{-}value$ for bootstrap linearity test. -1 if the test is not used
\item[] \texttt{e(grid)}	The number of grid points used
\item[] \texttt{e(trim)} The trim rate for grid search
\item[] \texttt{e(bs)} The number of bootstrapping
\end{itemize}

\noindent Macros
\begin{itemize}
\item[] \texttt{e(zx)}	The name of instrumental variables
\item[] \texttt{e(qx)}	The name of the threshold variable
\item[] \texttt{e(depvar)}	The name of the dependent variable
\item[] \texttt{e(indepvar)}	The name of the independent variable(s)
\item[] \texttt{e(properties)} The name of coefficient matrix and covariance matrix
\end{itemize}

\noindent Matrices
\begin{itemize}
\item[] \texttt{e(b)}	Estimates of coefficients
\item[] \texttt{e(V)}	The estimate of the covariance matrix
\item[] \texttt{e(CI)} $95\%$ asymptotic confidence interval for \texttt{b}
\end{itemize}

\section{Monte Carlo experiments}
In this section we illustrate the finite sample performance of the bootstrap linearity test. Some simulations for estimation were performed in Seo and Shin (2016). Here the model under $\mathcal{H}_0$ is linear, i.e. $\delta = 0$. We consider the following data generating process. Specific values of coefficients are different across simulations.
\[
\begin{split}
y_{i, t} &= \beta_1 y_{i, t-1} + \beta_2 x_{i, t} + (\delta_0 + \delta_1 y_{i, t-1} + \delta_2 x_{i, t}) 
\mathbbm{1}_{(x_{i, t} > 0)} + \varepsilon_{i, t}\\
\end{split} 
\]
We summarize more details of our simulation design in the following table.
\begin{table}[H]
\begin{center}
\begin{tabular}
{c|c|c}\hline
Parameter & Definition & Value \\\hline
$N$ & Cross-sectional sample size & 500 \\
$T$ & Time periods & 12 \\
$\#_{grid}$ & Number of grid points & 100 \\
$\#_{iter}$ & Number of iterations & 500 \\
$\alpha$ & Significance Level & 0.05 \\\hline
\end{tabular}
\end{center}
\end{table}
Moreover, $x_{i, t}$ and $\varepsilon_{i, t}$ were drawn independently from the centered normal distribution with standard deviation 1 and 0.25 respectively. For each iteration, we calculate bootstrap $sup \, W^{\ast }$ once following
e.g. Giacomini et al. (2013). Consequently, we
obtain 500 simulated $sup \, W$ and $sup \, W^{\ast }$ statistics. With this we
compute the bootstrap critical value, which is the empirical $\left(
1-\alpha \right) 100$-percentile of those 500 $sup \, W^{\ast }$ statistics, and
the rejection probability for the given $\alpha $, which is the proportion
of 500 $sup \, W$ statistics bigger than the bootstrap critical value.

\subsection{Test size}
Here we impose $(\beta_1, \beta_2, \delta_0, \delta_1, \delta_2) = (0.5, 0.8, 0.0, 0.0, 0.0)$  so that $\mathcal{H}_0: \delta = 0$ holds. This implies the true underlying model is linear. The simulated rejection probability was $0.066$ which is reasonably close to the true $\alpha=0.05$. Empirical distributions of $sup \, W$ and $sup \, W^*$ are as follow.
\begin{figure}[H]
\centering
\includegraphics[width=0.9\textwidth]{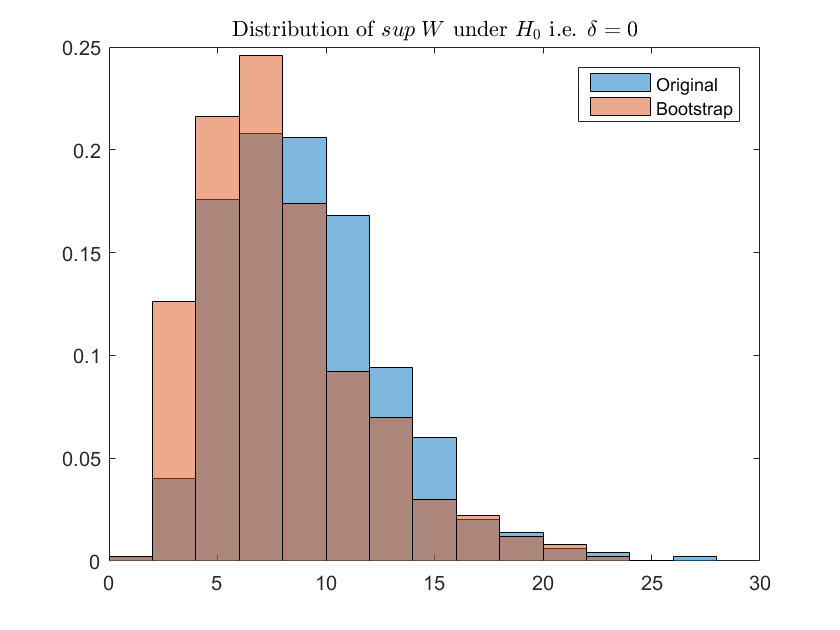}
\caption{Original and bootstrapped $sup\textit{-}Wald$ statistics under $\mathcal{H}_0$}
\end{figure}

\subsection{Test power}
Here we tested three sets of coefficient choices, maintaining $\mathcal{H}_1: \delta \neq 0$ holds. 
\begin{table}[H]
\begin{center}
\begin{tabular}
{c|c|c}\hline
Parameters & Values & Rejection Probability \\\hline
 & (0.5, 0.8, 0.0, -0.5, 0.0) & 1.00 \\
$(\beta_1, \beta_2, \delta_0, \delta_1, \delta_2)$ & (0.5, 0.0, 0.0, -0.5, 0.0) & 0.54 \\
 & (0.5, 0.0, 0.0, -0.9, 0.0) & 0.96 \\\hline
\end{tabular}
\end{center}
\end{table}

We observe that our test has significant power to reject $\mathcal{H}_0$ when $\mathcal{H}_1$ is true, especially when the true $\delta$ is sufficiently far from zero.

\section{Application}

We apply our method to evaluate the effect of obesity on worker's
productivity. Obesity is measured with Body Mass Index (BMI), weight in
kilograms divided by height in meters squared. Individuals whose BMI between
25 and 30 are considered to be overweight, and BMI of 30 or higher are
treated as obese. Using data from the British Cohort Study and the methods
described in the earlier section, we examine how BMI is associated with work
hours. For more detailed discussion, see Kim (2019).

In this example, we consider work hours and BMI of male workers using the
following model with a kink in BMI. 
\begin{equation*}
y_{it}=\beta _{0}+x_{it}\beta _{1}+q_{it}\beta _{2}+\delta (q_{it}-\gamma
)1\{q_{it}\geq \gamma \}+\alpha _{i}+\varepsilon _{it}, 
\end{equation*}%
where we present work hours as $y_{it}$ for an individual $i$ for a period $%
t,$ $x_{it}$ is family size and $q_{it}$ is BMI. We have two period panel
data $(t=1,2)$ and take the first difference as follows to remove $\alpha
_{i},$ the individual time-invariant charateristics that are associated with
work hours. 
\begin{equation*}
\Delta y_{i2}=\Delta x_{i2}\beta _{1}+\Delta q_{i2}\beta _{2}+\delta
(q_{i2}-\gamma )1\{q_{i2}\geq \gamma \}-(q_{i1}-\gamma )\delta 1\{q_{i1}\geq
\gamma \}+\Delta \varepsilon _{i2}. 
\end{equation*}%
To implement GMM estimation, we use four instrumental variables, birth
weight (bweight) and a worker's own childhood BMI (bmic) and parents' BMI
(bmim, bmid), for BMI variables of $\Delta q_{i2},q_{i2},q_{i1}$ in the first
differenced model.

After loading data, we first need to declare that the data is panel. The
default model for \texttt{xthenreg }is a dynamic model. Since we consider a
static model, not a dynamic model, we use \texttt{static} option. We also
impose a kink in the model by using \texttt{kink} option.

\begin{quotation}
\noindent \texttt{. use hour, clear}\newline
\texttt{. xtset ilabel time}\newline
\texttt{. xthenreg hour bmi hsize, endo(bmi) inst({\small bweight
		\linebreak \TEXTsymbol{>} \ bmic bmim bmid hsize})  kink static}\newline
\newline
\texttt{N = 768, T = 2}\newline
\texttt{Panel Var. = ilabel}\newline
\texttt{Time Var. = time}\newline
\texttt{Number of moment conditions = 7}
\end{quotation}

\begin{center}
\begin{tabular}{r|rrrrrr}
\hline
\texttt{\small hour} & \texttt{\small Coef.} & \texttt{\small Std.~Err.} & 
\texttt{\small z~} & \texttt{{\small P \TEXTsymbol{>}\TEXTsymbol{\vert}z%
\TEXTsymbol{\vert}}} & \texttt{\small [95\% Conf.} & \texttt{\small Interval]%
} \\ \hline
\texttt{\small hsize\_b} & \texttt{.4929009} & \texttt{.2707888} & \texttt{%
1.82} & \texttt{0.069} & \texttt{-.0378354} & \texttt{1.023637} \\ 
\texttt{\small bmi\_b} & \texttt{-.7547926} & \texttt{\ .9564381} & \texttt{%
-0.79} & \texttt{0.430} & \texttt{-2.629377} & \texttt{1.119792} \\ 
\texttt{\small kink\_slope} & \texttt{\ 2.533467} & \texttt{\ 2.170619} & 
\texttt{1.17} & \texttt{0.243} & \texttt{-1.720868} & \texttt{6.787802} \\ 
\texttt{\small r} & \texttt{29.04343} & \texttt{4.703686} & \texttt{6.17} & 
\texttt{0.000} & \texttt{\ 19.82437} & \texttt{38.26248} \\ \hline
\end{tabular}
\end{center}

\bigskip

The information preceding the table is as follows. \texttt{N} is the total
number of unique subjects, \texttt{T} is the number of time periods. \texttt{%
Number of moment conditions} is provided based on the choice of instruments.
In this example, we can obtain the same results by collecting all exogenous
variables into one place with \texttt{exo} option as follows.

\begin{quotation}
{ \noindent \texttt{. xthenreg hour bmi, endo(bmi) exo(hsize)
inst(bweight 
	\linebreak \TEXTsymbol{>} \ bmic  bmim bmid) kink static} }
\end{quotation}


We can also change the set of included and excluded instruments using the 
\texttt{inst} option. The number of moment conditions varies accordingly.

\begin{quotation}
{\noindent \texttt{. xthenreg hour bmi hsize, endo(bmi)
inst(bweight  
	\linebreak \TEXTsymbol{>} \ bmic  bmim bmid) kink static} \linebreak }

\noindent \texttt{N = 768, T = 2}\newline
\texttt{Panel Var. = ilabel}\newline
\texttt{Time Var. = time}\newline
\texttt{Number of moment conditions = 6}
\end{quotation}

\begin{center}
\begin{tabular}{r|rrrrrr}
\hline
\texttt{\small hour} & \texttt{\small Coef.} & \texttt{\small Std.Err.} & 
\texttt{\small z~} & \texttt{{\small P \TEXTsymbol{>}\TEXTsymbol{\vert}z%
\TEXTsymbol{\vert}}} & \texttt{\small [95\% Conf.} & \texttt{\small Interval]%
} \\ \hline
\texttt{\small hsize\_b} & .4964864 & .2774487 & \texttt{\small 1.79} & 
\texttt{\small 0.074} & -.0473031 & 1.040276 \\ 
\texttt{\small bmi\_b} & -.7783025 & 1.147452 & \texttt{\small -0.68} & 
\texttt{\small 0.498} & -3.027267 & 1.470662 \\ 
\texttt{\small kink\_slope} & 2.527627 & 2.29367 & \texttt{\small 1.10} & 
\texttt{\small 0.270} & -1.967884 & 7.023139 \\ 
\texttt{\small r} & 28.9816 & 6.002713 & \texttt{\small 4.83} & \texttt%
{\small 0.000} & 17.2165 & 40.7467 \\ \hline
\end{tabular}
\end{center}

\bigskip

We can estimate the model with a restriction on the sample.

\begin{quotation}
{\noindent \texttt{. xthenreg hour bmi if region==1, endo(bmi)
exo(hsize)
	\linebreak \TEXTsymbol{>} \  inst(bweight bmic bmim bmid) kink
static} \linebreak }

\noindent \texttt{N = 637, T = 2}\newline
\texttt{Panel Var. = ilabel}\newline
\texttt{Time Var. = time}\newline
\texttt{Number of moment conditions = 7}
\end{quotation}

\begin{center}
\begin{tabular}{r|rrrrrr}
\hline
\texttt{\small hour} & \texttt{\small Coef.} & \texttt{\small Std.Err.} & 
\texttt{\small z~} & \texttt{{\small P \TEXTsymbol{>}\TEXTsymbol{\vert}z%
\TEXTsymbol{\vert}}} & \texttt{\small [95\% Conf.} & \texttt{\small Interval]%
} \\ \hline
\texttt{\small bmi\_b} & -.3205365 & 1.841899 & \texttt{\small -0.17} & 
\texttt{\small 0.862} & -3.930593 & 3.28952 \\ 
\texttt{\small hsize\_b} & .5270078 & .3784874 & \texttt{\small 1.39} & 
\texttt{\small 0.164} & -.2148138 & 1.268829 \\ 
\texttt{\small kink\_slope} & 2.444602 & 7.610258 & \texttt{\small 0.32} & 
\texttt{\small 0.748} & -12.47123 & 17.36043 \\ 
\texttt{\small r} & 29.14014 & 17.36101 & \texttt{\small 1.68} & \texttt%
{\small 0.093} & -4.886813 & 63.16709 \\ \hline
\end{tabular}
\end{center}

\bigskip

Next we consider discontinuity in BMI effect without imposing a kink in the
model.%
\begin{equation*}
y_{it}=\beta _{0}+x_{it}\beta _{1}+q_{it}\beta _{2}+(\delta
_{0}+x_{it}\delta _{1}+q_{it}\delta _{2})1\{q_{it}>\gamma \}+\alpha
_{i}+\varepsilon _{it}. 
\end{equation*}%
By taking first difference, we obtain the following model and estimate it
with only \texttt{static }option.%
\begin{equation*}
\Delta y_{i2}=\Delta x_{i2}\beta _{1}+\Delta q_{i2}\beta _{2}+(\delta
_{0}+x_{i2}\delta _{1}+q_{i2}\delta _{2})1\{q_{i2}>\gamma \}-(\delta
_{0}+x_{i1}\delta _{1}+q_{i1}\delta _{2})1\{q_{i1}>\gamma \}+\Delta
\varepsilon _{i2}. 
\end{equation*}

\begin{quotation}
{\noindent \texttt{. xthenreg hour bmi, endo(bmi) exo(hsize)
inst(bweight
	\linebreak \TEXTsymbol{>} \  bmic bmim bmid) static} \linebreak }

\noindent \texttt{N = 768, T = 2}\newline
\texttt{Panel Var. = ilabel}\newline
\texttt{Time Var. = time}\newline
\texttt{Number of moment conditions = 7}
\end{quotation}

\begin{center}
\begin{tabular}{r|rrrrrr}
\hline
\texttt{\small hour} & \texttt{\small Coef.} & \texttt{\small Std.Err.} & 
\texttt{\small z~} & \texttt{{\small P \TEXTsymbol{>}\TEXTsymbol{\vert}z%
\TEXTsymbol{\vert}}} & \texttt{\small [95\% Conf.} & \texttt{\small Interval]%
} \\ \hline
\texttt{\small bmi\_b} & 6.069513 & 19.04971 & \texttt{\small 0.32} & \texttt%
{\small 0.750} & -31.26724 & 43.40626 \\ 
\texttt{\small hsize\_b} & -2.093733 & 16.48417 & \texttt{\small -0.13} & 
\texttt{\small 0.899} & -34.40211 & 30.21464 \\ 
\texttt{\small cons\_d} & 106.3078 & 626.0473 & \texttt{\small 0.17} & 
\texttt{\small 0.865} & -1120.722 & 1333.338 \\ 
\texttt{\small bmi\_d} & -5.569615 & 27.32028 & -0.20 & 0.838 & -59.11639 & 
47.97716 \\ 
\texttt{\small hsize\_d} & 4.898759 & 16.62541 & 0.29 & 0.768 & -27.68644 & 
37.48395 \\ 
\texttt{\small r} & 25.64312 & 14.89532 & \texttt{\small 1.72} & \texttt%
{\small 0.085} & -3.551176 & 54.83741 \\ \hline
\end{tabular}
\end{center}

\end{document}